\documentclass[useAMS,usenatbib,usegraphicx]{mn2e}
 
\title
{The contribution of star-forming galaxies to the cosmic radio background}

\author[P.~P.~Ponente et~al.]
{
P.~P.~Ponente$^{1,2}$, Y.~Ascasibar$^3$, and J.~M.~Diego$^{1}$\\
$^1$\,IFCA, Instituto de F\'\i sica de Cantabria (UC-CSIC), Av. de Los Castros s/n, 39005 Santander, Spain\\
$^2$\,Departamento de F\'\i sica Moderna, Universidad de Cantabria. Av. de Los Castros s/n, 39005 Santander, Spain\\ 
$^3$\,Departamento de F\'{i}sica Te\'{o}rica, Universidad Aut\'{o}noma de Madrid, Madrid 28049, Spain
}

\date{{\bf Final version} (\today)}

\newcommand{\aap}{A\&A}
\newcommand{\aaps}{A\&AS}
\newcommand{\aj}{AJ}
\newcommand{\apj}{ApJ}
\newcommand{\apjl}{ApJ}
\newcommand{\apjs}{ApJS}

\newcommand{\araa}{ARA\&A}
\newcommand{\mnras}{MNRAS}
\newcommand{\nat}{Nature}

\newcommand{\lsim} {\mathrel{\hbox{\rlap{\lower.55ex \hbox{$\sim$}}
                             \kern-.3em \raise.4ex \hbox{$<$}}}} 
\newcommand{\msun}{{\rm M}_\odot}
\newcommand{\arcade}{{\sc arcade2 }}
\newcommand{\sfr}{\dot{\rho}_*}
 
\newcommand{\ff}{_{\rm ff}}
\newcommand{\syn}{_{\rm syn}}
\newcommand{\ir}{_{\rm IR}}
\newcommand{\dd}{{\rm d}}

\usepackage{color}
\newcommand{\change}{}

\begin{document} 
 
\maketitle 
 
\begin{abstract}
Recent measurements of the temperature of the sky in the radio band, combined with literature data, have convincingly shown the existence of a cosmic radio background with an amplitude of $\sim 1$~K at 1~GHz and a spectral energy distribution that is well described by a power law with index $\alpha \simeq -0.6$.
The origin of this signal remains elusive, and it has been speculated that it could be dominated by the contribution of star-forming galaxies at high redshift \change{if the far infrared-radio correlation $q(z)$ evolved} in time.
\change{We fit observational data from several different experiments by the relation
$q(z) \simeq q_0 - \beta \log(1+z)$
with $q_0 = 2.783 \pm 0.024$ and $\beta = 0.705 \pm 0.081$ and estimate the total radio emission of the whole galaxy population at any given redshift from the cosmic star formation rate density at that redshift.
It is found that} star-forming galaxies can only account for $\sim$13 percent of the observed intensity of the cosmic radio background.
\end{abstract}


\section{Introduction}

Although the detection of diffuse radio emission dates back to \citet{Jansky33}, the origin of the cosmic radio background (CRB) is still a mystery.
The recent data obtained by the Absolute Radiometer for Cosmology, Astrophysics and Diffuse Emission ({\sc arcade\,2}) has revived the interest in this question, detecting a diffuse background at frequencies between 3 and 10~GHz that is more then 5$\,\sigma$ above the COBE/FIRAS measurement of the temperature of the cosmic microwave background (CMB) and well in excess of current estimates based on radio source counts.
More precisely \citep{arcade}, the inferred value of the antenna temperature as a function of frequency can be expressed as
\begin{equation}
T(\nu) = \frac{ h\nu / k }{ \exp(h\nu/kT_{\rm CMB}) - 1 } + T_{\rm R} \left( \frac{\nu}{\nu_0} \right)^{\alpha-2}
\end{equation}
where $T_{\rm CMB} = 2.729 \pm 0.004$~K denotes the thermodynamic temperature of the CMB, $T_{\rm R} = 1.19 \pm 0.14$~K is the normalization of the radio background at $\nu_0 = 1$~GHz, and $\alpha=-0.62\pm0.04$ is the spectral index of the CRB, consistent with synchrotron emission from normal galaxies \citep[see e.g.][]{Condon_92}.

The observed emission is most likely of extragalactic origin \citep{Kogut+11}, and several candidates have been considered by \citet{Singal+10}.
Radio source counts detected by current surveys, sensitive to flux densities above $S_{\rm 1.4\,GHz} \ga 10~\mu$Jy, cannot explain more than $\sim 10$ per cent of the signal \citep[][]{Gervasi+08,Massardi+10,Vernstrom_et_al_2011}, and low-surface brightness sources missed by these surveys may contribute, at most, an additional $15$ per cent.
Diffuse emission in regions far from galaxies is ruled out due to the overproduction of X-rays and $\gamma$-rays, so the only possible explanation is that the cosmic radio background is dominated by faint sources below the threshold of $10~\mu$Jy \citep{Singal+10}.

According to \citet{Singal+10}, radio supernovae make a negligible contribution, and radio-quiet quasars may be responsible for only a few per cent of the emission.
Thermal bremsstrahlung from the hot gas in galaxy clusters has been shown to contribute about $0.01-0.02$~K at $\nu=1$~GHz (see e.g. \citealp{Ponente_et_al_2011}), and the 
most reasonable candidate to explain the bulk of the CRB seems to be the population of ordinary star-forming galaxies at high redshift.
 
Some authors \citep[e.g.][]{Oh1999,Cooray_Furlanetto_2004} have tried to
estimate the contribution of free-free emission from 
star-forming galaxies to the radio background by resorting to
phenomenological prescriptions to relate halo mass and star formation
activity at different redshifts. \\
However, if the far infrared-radio correlation (FRC) observed for local galaxies holds at all redshifts, there must be a tight relation between the radio and infrared backgrounds \citep{HaarsmaPartridge98,DwekBarker02}.
From the measured intensity of the latter, one concludes that the contribution of star-forming galaxies must be of the order of $5-10$ per cent.

During the last years, the advances in infrared and sub-millimetric instrumentation have made it possible to investigate the evolution of the FRC over a large fraction of the age of the Universe, and several recent studies \citep[e.g.][]{Ivison_et_al_2010_a,Ivison_et_al_2010_b,Michalowski} suggest that the correlation is linear at all times, but the normalization is offset towards increasing radio loudness at high redshifts, boosting the expected signal from star-forming galaxies by a significant amount.
In the present work, we make a quantitative estimate of the contribution of star-forming galaxies to the CRB.
The prescription followed to assign radio luminosities as a function of the instantaneous star formation rate is detailed in Section~\ref{sec_individual}.
The evolution of the far infrared-radio correlation is discussed in Section~\ref{sec_FRC}, and the implications for the cosmic radio background are shown in Section~\ref{sec_CRB}.
Our main conclusions are briefly summarized in Section~\ref{sec_conclus}.

\section{Radio emission from individual galaxies}
\label{sec_individual}

In normal galaxies, radio emission is always associated to star formation \citep[see e.g.][]{Condon_92}.
Young, massive stars produce intense ultraviolet radiation that ionizes the surrounding medium, and thermal bremsstrahlung from these free electrons (often referred to in the radio literature as \emph{free-free} emission) makes a significant contribution to the galaxy spectra in the few-GHz range.
On the other hand, stars with $M > 8~\msun$ explode as Type II and Type Ib supernovae at the end of their life cycle.
Supernova remnants are thought to accelerate most of the relativistic electrons in normal galaxies, and they constitute the main source of the synchrotron emission that dominates at low frequencies.

Assuming a pure Hydrogen plasma with an electron temperature $T_{\rm e} \sim 10^4$~K, the free-free luminosity of a galaxy is approximately given by
\begin{equation}
\frac{ L\ff }{\rm 3.2 \times 10^{-39}~erg~s^{-1}~Hz^{-1} }
\approx\!
\left(\! \frac{ \nu       }{ \rm GHz     } \!\right)^{\!\!-0.1}\!\!
\left(\! \frac{ n_{\rm e} }{ \rm cm^{-3} } \!\right)^{\!\!2}\!
\left(\! \frac{ V_{\rm e} }{ \rm cm^{3}  } \!\right)
\end{equation}
where $\nu$ denotes the photon frequency, $n_{\rm e}$ is a characteristic electron density, and $V_{\rm e}$ represents the total volume occupied by the radio-emitting, ionized H{\sc ii} regions \citep{Rubin68,Oh1999}.
This volume is set by the condition that the number of ionizing photons $Q$ emitted by the stars per unit time is equal to the recombination rate
\begin{equation}
Q = n_{\rm e}^2\ \alpha_{\rm B}\ V_{\rm e}
\label{eq_ionizing}
\end{equation}
with $\alpha_{\rm B} = 2.6 \times 10^{-13}~{\rm cm^{3}~s^{-1}}$ being appropriate for case-B recombination at $T_{\rm e} \sim 10^4$~K.
According to stellar population synthesis models \citep[e.g.][]{Leitherer_95, Molla_et_al_2009},
\begin{equation}
\frac{ Q }{ \rm 1.5 \times 10^{53}~s^{-1} } \approx \frac{ \Psi }{ \rm \msun~yr^{-1} }
\label{eq_Q}
\end{equation}
where $\Psi$ is the current star formation rate (SFR) of the galaxy, assuming a \citet{Salpeter_1955} initial mass function (IMF) between $0.1$ and $100~\msun$.
In the end, the predicted free-free luminosity
\begin{equation}
\frac{ L\ff }{\rm 1.8 \times 10^{27}~erg~s^{-1}~Hz^{-1} }
\approx
\frac{ \Psi }{ \rm \msun~yr^{-1} }
\left( \frac{ \nu }{ \rm GHz } \right)^{\!\!-0.1}
\label{eq_ff}
\end{equation}
scales roughly proportionally with the instantaneous star formation rate.

Computing the synchrotron luminosity from first physical principles is much more involved, since it requires knowledge of the amount of cosmic rays injected by supernovae, their spectrum, and the conditions of the surrounding medium (most notably, its density structure and the intensity of the magnetic field).
Observationally \citep{Condon_92}, non-thermal synchrotron emission is about 10 times more luminous than the free-free continuum at $\nu=1$~GHz, and its spectral index is close to $\sim 0.7$ for a broad range of star-forming galaxies.
In addition, there is a tight correlation between the synchrotron luminosity and the thermal radiation emitted by the dust in the infrared, which is powered by the stellar ultraviolet light and is thus another tracer of the star formation rate.
The observed far infrared-radio correlation suggests \citep[but see e.g.][for a different point of view]{Lacki_et_al_2010,Lacki_Thompson} that synchrotron emission is also proportional to the SFR, implying that
\begin{equation}
\frac{ L\syn }{\rm 1.8 \times 10^{28}~erg~s^{-1}~Hz^{-1} }
\approx
\frac{ \Psi }{ \rm \msun~yr^{-1} }
\left( \frac{ \nu }{ \rm GHz } \right)^{\!\!-0.7}.
\label{eq_syn_0}
\end{equation}

\section{Evolution of the FRC}
\label{sec_FRC}

Since most of the contribution of normal galaxies to the cosmic radio background observed today is due to their synchrotron emission, with thermal bremsstrahlung \citep[see e.g.][]{Oh1999, Seiffert+11,Ponente_et_al_2011} providing only a minor correction at the level of a few percent, equation~(\ref{eq_syn_0}) has a crucial importance.
In particular, the intensity of the CRB is extremely sensitive to the evolution in time of the relation between SFR and radio luminosity.

It is not clear, though, whether the far infrared-radio correlation should evolve with redshift, and current observational evidence is far from being conclusive.
While several recent studies \citep[e.g.][]{Ibar_et_al_2008,Sargent_et_al_2010} are consistent with no evolution in the FRC, some others report systematic trends with redshift \citep[e.g.][]{Vlahakis_et_al_2007,Beswick_et_al_2008,Seymour_et_al_2009,Michalowski}.

The main source of uncertainty is that normal galaxies are rather faint in the radio band.
According to equation~(\ref{eq_syn_0}), only the most intense starbursts, with instantaneous SFR in excess of $\Psi \ga 30~\msun~yr^{-1}$, would be detectable at $z \ge 1$ by current surveys, whose sensitivity at 1.4~GHz is of the order of $\sim 10~\mu$Jy.

\begin{figure}
\centering
\includegraphics[width=8.5cm]{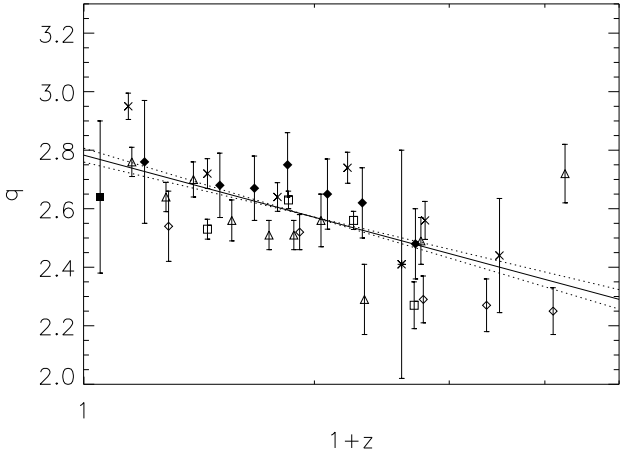}
\caption{Evolution of the far infrared-radio correlation.
Least-squares fit (solid line) with $1\sigma$ limits (dash lines) to the observational data from \citep[full squares]{Bell03}, \citep[stars]{Murphy+09}, \citep[open diamonds]{Michalowski}, \citep[triangles]{Sargent_et_al_2010}, \citep[full diamonds]{Bourne_et_al_2011}, \citep[crosses]{Ivison_et_al_2010_a} and \citep[open squares]{Ivison_et_al_2010_b}.}
\label{fig_q}
\end{figure}

One possible solution \citep[see e.g.][]{Marsden_et_al_2009,Pascale_et_al_2009,Patanchon_et_al_2009} is to stack the confusion-limited and sensitivity-limited radio images at the positions of thousands of infrared-selected galaxies.
In doing so, one increases the signal-to-noise ratio and reduces the contribution of radio-loud active galactic nuclei (AGN), probing a population that is more representative of normal galaxies.
This procedure has been applied by \citet{Ivison_et_al_2010_a} to a mid infrared-selected sample of galaxies, obtaining that
\begin{equation}
q \equiv \log\frac{ L\ir \,/\, {\rm 3.75\times 10^{12}~W} }{ L_{\rm 1.4\, GHz} \,/\, {\rm W~Hz^{-1}} } \propto (1+z)^{\gamma}
\end{equation}
with $\gamma = -0.15 \pm 0.03$.
Both the total infrared luminosity $L\ir$ (defined from 8 to $1000~\mu$m)\footnote{The difference with e.g. the far-infrared band (from 60 to $100~\mu$m) is about a factor of two. and the radio power $L_{\rm 1.4\, GHz}$ are given at the rest-frame of the source, using a k-correction based on spectral templates.}
A similar analysis \citep{Ivison_et_al_2010_b} is consistent with no evolution, $\gamma = -0.04 \pm 0.03$, but discarding the least reliable data at $z < 0.5$ yields $\gamma = - 0.26 \pm 0.07$.

Alternatively, one may detect high-redshift star-forming galaxies by observing their rest-frame infrared dust emission, shifted towards sub-millimeter wavelengths.
Based on a sample of 76 sub-millimeter galaxies with measurements in the radio band, \citet{Michalowski} conclude, in agreement with previous studies \citep[e.g.][]{Kovacs+06,Murphy09} that the radio emission of high-redshift galaxies scales linearly with the SFR, but the normalization is about a factor of two higher than for local samples.

Although selection effects \change{\citep[see e.g.][]{Sargent_et_al_2010} and potential biases arising from spectral templates \citep{Bourne_et_al_2011} cannot be completely excluded, a combination of different data sets} is fairly well reproduced by
\begin{equation}
 q(z) = q_0 - \beta \log(1+z)
\label{eq_q}
\end{equation}
with $q_0 = 2.783 \pm 0.024$ and $\beta = 0.705 \pm 0.081 $ (see Figure~\ref{fig_q}).
Assuming that $L\ir \propto \Psi$ and that the constant of proportionality does not vary with redshift, this implies that the synchrotron luminosity of a given galaxy scales as
\begin{equation}
\frac{ L\syn }{\rm 1.8 \times 10^{28}~erg~s^{-1}~Hz^{-1} }
\approx
\frac{ \Psi }{ \rm \msun~yr^{-1} }
\left( \frac{ \nu }{ \rm GHz } \right)^{\!\!-0.7}
\!\!(1+z)^\beta
\label{eq_syn_1}
\end{equation}
In other words, we assume that the infrared luminosity is an unbiased tracer of the SFR and that all the evolution of the FRC is due to the conversion between SFR and radio luminosity.

\section{The cosmic radio background}
\label{sec_CRB}

The specific intensity of the cosmic background at any given frequency is given by the integral along the line of sight
\begin{equation}
I_\nu = \frac{ c }{ 4\pi H_0 } \int_0^\infty \frac{ \epsilon_{\nu'}(z) }{ (1+z)\, E(z) }\ \dd z
\label{eq_intensity}
\end{equation}
of the average emissivity per unit volume $\epsilon_{\nu'}$.
In this formula, $c$ and $H_0$ denote the speed of light and the Hubble constant, respectively,
\begin{equation}
E(z) = \sqrt{ \Omega_{\rm m}(1+z)^3 + \Omega_{\rm k}(1+z)^2 + \Omega_\Lambda }
\end{equation}
reflects the cosmological expansion, and $\nu' = \nu (1+z)$ is the initial frequency at which the photons observed today with a frequency $\nu$ were emitted.
We adopt a WMAP7 (seven-year observation) cosmology with $\Omega_{\rm m} = 0.27$, $\Omega_{\rm k} = 0$, $\Omega_\Lambda = 0.73$, and $H_0 = 71$~km~s$^{-1}$~Mpc$^{-1}$ \citep{WMAP7} and compute the brightness temperature of the CRB as
\begin{equation}
T(\nu) = \frac{ c^2\, I_\nu }{ 2k\nu^2 }
\label{eq_temp_conversion}
\end{equation}
using the Rayleigh-Jeans approximation, where $k$ is the Boltzmann constant.

By definition, the average emissivity at a given redshift is the sum
\begin{equation}
\epsilon_{\nu'}(z) = \int_0^\infty n(\Psi,z)\ L_{\nu'}(\Psi)\ \dd\Psi
\end{equation}
of the contributions of all the galaxies at that redshift, with $n(\Psi,z)$ representing the number density of galaxies with SFR between $\Psi$ and $\Psi+\dd\Psi$ at redshift $z$.
As long as the relation between luminosity and instantaneous star formation rate is linear, $L_{\nu'} = \kappa(\nu',z)\,\Psi$, as indicated by equations~(\ref{eq_ff}) and~(\ref{eq_syn_1}), one can express the total emissivity
\begin{equation}
\epsilon_{\nu'}(z)
= \kappa(\nu',z) \int_0^\infty n(\Psi,z)\ \Psi\ \dd\Psi
= \kappa(\nu',z)\ \sfr(z)
\label{eq_emissivity}
\end{equation}
in terms of the cosmic SFR density $\sfr$ \citep{DwekBarker02}.
The emissivity of the free-free and synchrotron components can be taken into account simultaneously as
\begin{equation}
\epsilon_{\nu'}(z) = \left[ \kappa\ff(\nu',z) + \kappa\syn(\nu',z) \right]\ \sfr(z)
\end{equation}
with
\begin{equation}
\frac{ \kappa\ff(\nu',z) }{\rm 1.8 \times 10^{27}\ erg\,s^{-1}\,Hz^{-1}\,\msun^{-1}\,yr }
=
\left( \frac{ \nu' }{ \rm GHz } \right)^{\!\!-0.1}
\label{eq_kappa_ff}
\end{equation}
and
\begin{equation}
\frac{ \kappa\syn(\nu',z) }{\rm 1.8 \times 10^{28}\ erg\,s^{-1}\,Hz^{-1}\,\msun^{-1}\,yr }
=
\left( \frac{ \nu' }{ \rm GHz } \right)^{\!\!-0.7}
\!\!\!(1+z)^\beta
\label{eq_kappa_syn}
\end{equation}
where $\beta=0$ for a non-evolving far infrared-radio correlation, and $\beta=0.705 \pm 0.0801$ to fit the \change{data plotted in Figure~\ref{fig_q}}.

\begin{figure}
\centering
\includegraphics[width=8.0cm]{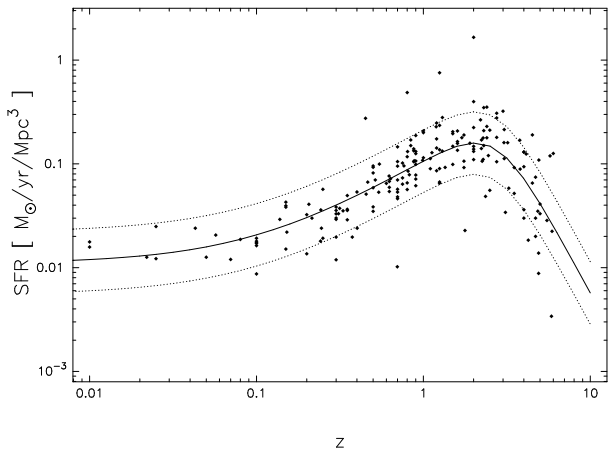}
\caption{Cosmic star formation history.
The solid line shows the best fit provided by expression~(\ref{eq_sfr}) to the data points compiled by \citet{Michalowski}, and dotted lines illustrate an uncertainty of a factor of 2.}
\label{fig_SFR}
\end{figure}

The evolution of the cosmic SFR density has been extensively studied during the last decade, and several compilations of observational data exist in the literature \citep[e.g.][]{Somerville_et_al_2001,Ascasibar_et_al_2002,Hopkins_2004,Hopkins_Beacom_2006,Michalowski}.
In the present work, we have adopted the parametrization of \cite{Cole_et_al_2001}
\begin{equation}
\frac{ \sfr(z) }{\rm \msun~yr^{-1}~Mpc^{-3} } = \frac{ a + bz }{ 1 + \left ( \frac{z}{c} \right)^d }
\label{eq_sfr}
\end{equation}
and fit the observational data points reported in Table~A.4 of \citet{Michalowski}\footnote{Conversion to a Salpeter IMF between $0.1$ and $100\msun$ and a WMAP7 cosmology \citep[following the prescription in][]{Ascasibar_et_al_2002} amounts to a negligible correction.}.
The optimal values $(\,a,b,c,d\,) = (\, 0.011,\, 0.097,\, 2.73,\, 3.96\,)$ have been obtained by means of the FiEstAS sampling technique \citep{Ascasibar_2008}, a Monte Carlo integration scheme based on the Field Estimator for Arbitrary Spaces \citep[FiEstAS;][]{AscasibarBinney_2005,Ascasibar_2010}.
The resulting cosmic star formation history is plotted as a solid line in Figure~\ref{fig_SFR}.

\begin{figure}
\centering
\includegraphics[width=8.5cm]{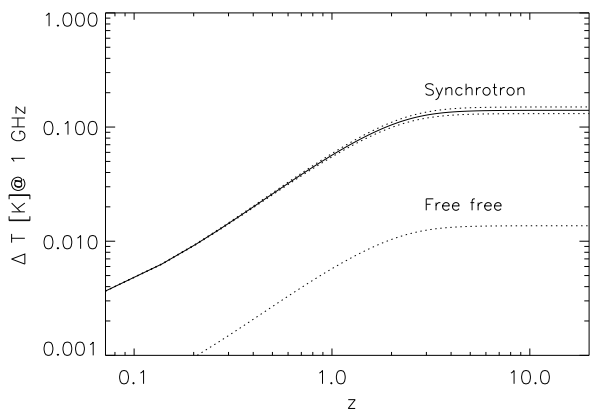}
\caption
{
\change
{
Integrated radio emission, observed at 1 GHz, from normal galaxies up to redshift $z$.
$\Delta T$ refers to the excess signal above the CMB temperature.
The solid line shows the contribution of synchrotron emission, assuming $\beta = 0.705$, and the errors associated to the least-squares fit ($\sim 0.01$ at the end of the integration) are indicated by the dashed lines.
The contribution of free-free emission is plotted as a dotted line.
}
}
\label{fig_integrated}
\end{figure}

Combining expressions~(\ref{eq_intensity}), (\ref{eq_temp_conversion}), (\ref{eq_emissivity}), (\ref{eq_kappa_ff}), and (\ref{eq_sfr}), we estimate that the contribution of free-free emission from star-forming galaxies to the cosmic radio background is
\begin{equation}
\frac{ T\ff }{ \rm 0.0137\ K } = \left( \frac{ \nu }{ \rm GHz } \right)^{\!\!-2.1}
\end{equation}
whereas, using expression~(\ref{eq_kappa_syn}), synchrotron emission yields
\begin{equation}
\frac{ T\syn }{ 0.0817\rm \ K } = \left( \frac{ \nu }{ \rm GHz } \right)^{\!\!-2.7}
\end{equation}
for $\beta=0$ and
\begin{equation}
\frac{ T\syn }{ \rm 0.1402\ K } = \left( \frac{ \nu }{ \rm GHz } \right)^{\!\!-2.7}
\end{equation}
for $\beta=0.705$.
\change
{
As can be seen in Figure~\ref{fig_integrated}, the signal is dominated by galaxies at $z<3$, due to the combined effects of distance dimming and the declining behavior of the cosmic star formation rate at high redshift.
}

\section{Discussion and Conclusions}
\label{sec_conclus}

In the present work, we have considered the relationship between the cosmic star formation rate and the radio background from \change{star-forming galaxies in the light of recent measurements of the far infrared-radio correlation at different redshifts}, attempting to give a further look into the significant missing flux that has been reported by the \arcade team.

Our main result is that normal galaxies can \emph{not} be responsible for the observed signal.

Although we think this conclusion is fairly robust, there is always some room for uncertainty.
Radio emission in local galaxies has been thoroughly studied, and its properties are well known \citep[see e.g.][]{Condon_92}, but different gas compositions and/or temperatures may affect the conversion factor between SFR and radio emission by a significant amount, of the order of several tens of percent.

On the other hand, we use the cosmic star formation rate density to constrain the average emissivity of the universe \citep{DwekBarker02}.
In contrast to radio source counts, where a population of faint objects below the detection threshold is very difficult to rule out \citep{Singal+10,Vernstrom_et_al_2011}, it would be extremely unlikely that our proposed fit underestimates the average SFR by more than a factor of two (dotted lines in Figure~\ref{fig_SFR}).
Uncertainties in the IMF cancel out with the production rate of ionizing photons given by expression~(\ref{eq_Q}) and are not expected to affect the present analysis significantly.

The most important source of uncertainty is the possible evolution of the far infrared-radio correlation.
Current observations seem to be compatible with $\beta = 0.705 \pm 0.081$, increasing the expected emission from normal galaxies by about 70 per cent with respect to the case of no evolution.
Using an extreme value $\beta = 1$ would boost the signal by only an additional 35 percent.

Nonetheless, it is worth noting that the high-redshift points in Figure~\ref{fig_q} \citep[e.g. in the samples of][]{Michalowski,Murphy+09} are dominated by sub-millimeter galaxies.
There is some discussion in the literature that these sources, whose contribution to the total SFR at $z \sim 1-2$ is only of the order of ten per cent \citep[see Figure 4 in][]{Michalowski}, may be radio-bright compared to normal galaxies and introduce some evolution in the observed FRC that does not apply to star formation as a whole\footnote{Just by removing the \citet{Michalowski} data from the least-squares fit, the best value of $\beta$ decreases to $0.57 \pm 0.093$. Furthermore, the data at $z<1$ are compatible with $\beta=0$ \citep[see the discussion in][]{Sargent_et_al_2010}.}.
In fact, one would expect on theoretical grounds that the FRC of normal galaxies evolved in the opposite direction ($\beta<0$).
On the one hand, star formation at $z\sim1$ is heavily obscured by dust, and the approximation that all the ultraviolet luminosity is re-radiated in the infrared is very good.
In the local universe, some fraction of the ionizing photons is able to escape, and the infrared luminosity per unit SFR should be lower.
On the other hand, galaxies at high redshift should produce less radio emission because the energy density of the CMB scales as $(1+z)^4$, and the relativistic electrons injected by supernovae lose more energy through inverse Compton scattering \citep[see e.g.][]{Carilli_Yun_1999,Carilli+2008,Murphy+09}.
Both effects, especially the latter, would only strengthen our conclusions, and the estimate with $\beta=0.705$ should arguably be regarded as an upper limit.

According to our results, radio emission from star-forming galaxies could explain up to $\sim 13$ per cent of the intensity of the CRB.
Even taking all the possible uncertainties into account, we are still far from the 1.19 Kelvin reported by \arcade at 1~GHz.
Although evolution of the FRC at $z<3$ has to be further investigated, current data strongly suggest that it only results in a relatively minor boost to the contribution of normal galaxies, and hence we can rule them out as the main source for the radio background.
As shown in Figure~\ref{fig_integrated}, the contribution of galaxies at higher redshifts is negligible.

Since relatively bright point sources, as well as Galactic or extragalactic diffuse emission have also been ruled out \citep[and references therein]{Singal+10}, there are few alternatives left to explain the observed cosmic radio background.
Some possibilities are:
\begin{enumerate}
\item The \arcade measurement is incorrect, or it is contaminated by Galactic foregrounds.
Being perfectly consistent with independent measurements at longer wavelengths \citep[e.g.][]{Haslam+82,deOliveira-Costa+08,RogersBowman08}, we think this possibility is unlikely.
\item Faint star-forming galaxies at high redshift are extremely radio bright, perhaps due to an enhanced magnetic field or AGN activity with respect to the brightest objects at that redshift \citep[the possibility favored by][]{Singal+10}.
\item There is a new population of numerous and faint radio sources waiting to be discovered.
\end{enumerate}

To sum up, the nature of the cosmic radio background poses an exciting challenge for radio astronomy, to be faced in the upcoming era of Expanded Very Large Array (EVLA) and Square Kilometre Array (SKA).

 \section*{Acknowledgments}

We would like to thank the anonymous referee for his/her constructive suggestions and an extremely careful reading of the manuscript.
Funding for the present work has been provided by the \emph{Ministerio de Ciencia e Innovaci\'on} (Spain) through projects AYA\,2010-21887-C04-03, AYA\,2010-21766-C03-01 and CSD2010-00064. PPP acknowledges support from the Spanish Ministerio de Ciencia e Innovaci\'on and CSIC for an I3P grant. JMD acknowledges support from the CSIC international project 2009-50I-198.



\end{document}